\documentclass[prl,reprint,superscriptaddress]{revtex4-1}

\usepackage{amssymb,amsmath,amsfonts,amsthm, amscd, mathrsfs,helvet, mathtools}

\usepackage{bm, stmaryrd}
\usepackage{graphicx,verbatim}
\usepackage[usenames, dvipsnames]{color}
\usepackage{psfrag}
\usepackage[all]{xy}

\theoremstyle{plain}

\newtheorem*{theorem*}{Theorem}

\newtheorem*{proposition*}{Proposition}

\newcommand{\tensore}[1]{{\bf \underline{#1}}}

\definecolor{brightBlue}{rgb}{0,0,1}

\definecolor{Violet}{rgb}{0.47,0,1}

 \DeclareMathOperator{\str}{Str}

\def\ha{\mbox{\small $\frac{1}{2}$}}

\def\1{\tensore{1}}
\def\2{\tensore{2}}
\def\3{\tensore{3}}
\def\4{\tensore{4}}

\def\beq{\begin{equation}}
\def\eeq{\end{equation}}

\def\be{\begin{equation}}
\def\ee{\end{equation}}

\def\beqz{\begin{equation*}}
\def\eeqz{\end{equation*}}
\def\bea{\begin{eqnarray}}
\def\eea{\end{eqnarray}}
\def\nn{\nonumber}

\def\f{\mathfrak{f}}

\newcommand{\coloneqq}{:=}

\newcounter{comm}
\newcounter{piq}
\newcounter{treqle}
\newcommand{\details}[1]
{
}

\def\dd{d}
\def\dt{\widetilde{d}}
\def\JJ{J}
\def\Jt{\widetilde{J}}

\def\AA{A}

\def\Lax{{\cal{L}}}

\def\Lag{{L}}

\def\Id{{\mathsf{1}}}

\def\eom{{\cal E}}
\def\Klimcik{{Klim$\check{\text{c}}$\'{\i}k} }
\def\mc{{\cal Z}}

\begin{document}

\title{An integrable deformation of the  $AdS_5 \times S^5$ superstring action}

\author{F. Delduc}
\affiliation{Laboratoire de Physique, ENS Lyon
et CNRS UMR 5672, Universit\'e de Lyon, \\
46, all\'ee d'Italie, 69364 LYON Cedex 07, France}
\author{M. Magro}
\affiliation{Laboratoire de Physique, ENS Lyon
et CNRS UMR 5672, Universit\'e de Lyon, \\
46, all\'ee d'Italie, 69364 LYON Cedex 07, France}
\author{B. Vicedo}
\affiliation{School of Physics, Astronomy and Mathematics,
University of Hertfordshire,\\
College Lane,
Hatfield AL10 9AB,
United Kingdom}

\begin{abstract}
An integrable deformation of the type IIB $AdS_5 \times S^5$ superstring action is
presented.
The deformed field equations, Lax connection, and $\kappa$-symmetry transformations are given. The original $\mathfrak{psu}(2,2|4)$ symmetry is expected to become $q$-deformed.

\end{abstract}
\maketitle

\section{Introduction}

Integrability plays a central role in the study of the AdS/CFT correspondence
\cite{Maldacena:1997re, *Gubser:1998bc,*Witten:1998qj}
between type IIB superstring theory on the $AdS_5 \times S^5$ background
\cite{Metsaev:1998it} and
the maximally supersymmetric Yang-Mills gauge theory in four
dimensions (see \cite{Beisert:2010jr} for a review). On the Anti-de Sitter
side of this correspondence, integrability entered the scene with the discovery
that the lagrangian field equations of the $AdS_5 \times S^5$ theory can be
recast in the zero curvature form \cite{Bena:2003wd}.
This implies the existence of an infinite number of conserved quantities.

It is quite natural to seek deformations of the $AdS_5 \times S^5$ superstring which preserve this integrable structure. An important example of such an integrable deformation is the so called $\beta$-deformation associated with strings on the Lunin-Maldacena background \cite{Lunin:2005jy}.
The integrability of this model was shown in \cite{Frolov:2005ty, Frolov:2005dj} 
(see also the review \cite{Zoubos:2010kh} and references therein). Here we shall take a more systematic approach to the construction of integrable deformations by demanding the deformed theory to be integrable from the very outset. This requires approaching the problem from the hamiltonian perspective.

Let us recall that in order to prove integrability in the hamiltonian formalism, one must show
the existence of an infinite number of conserved quantities in involution.
More precisely, this follows at once if the Poisson bracket of the
hamiltonian Lax matrix can be shown to take the specific form in \cite{Maillet:1985fn,Maillet:1985ek}.
This was achieved in the case of the $AdS_5 \times S^5$ superstring in \cite{Magro:2008dv}.

The algebraic structure underpinning this property of the $AdS_5 \times S^5$ superstring was
identified in \cite{Vicedo:2010qd}. By utilising this structure, an alternative Poisson
bracket with the same property was subsequently constructed in \cite{Delduc:2012vq}.
Moreover, this second Poisson bracket is compatible with
the original one, giving rise to a one-parameter family of Poisson brackets
sharing the same property \cite{Maillet:1985fn,Maillet:1985ek} which ensures integrability.

These features of the superstring theory are in fact shared with bosonic integrable $\sigma$-models \cite{Delduc:2012qb}. In this latter context, the two compatible Poisson brackets were used very recently in \cite{arxiv-1308.3581} as a building block for constructing
integrable $q$-deformations of the principal chiral model associated with a compact
Lie group and of the $\sigma$-model on a symmetric space $F/G$ with $F$ compact.
In the case of the principal chiral model, the deformation coincides with the Yang-Baxter $\sigma$-model introduced by \Klimcik in \cite{Klimcik:2002zj}. A key characteristic of this
procedure is that the integrability of the deformed theories is automatic since it is
used as an input in the construction. Moreover, an interesting output is that
the symmetry associated with left multiplication in the original models is
deformed into a classical $q$-deformed Poisson-Hopf algebra.

It is possible to generalize the method developed
in \cite{arxiv-1308.3581}  to deform the $AdS_5 \times S^5$ superstring theory.
The whole construction is carried out at the hamiltonian level and will
be presented in detail elsewhere. The purpose of this letter is to
present the deformed action and indicate its properties.

\section{Setting}

We begin by recalling the necessary ingredients for defining the
$AdS_5 \times S^5$ superstring
action (see \cite{Arutyunov:2009ga} for more details). Define the projectors $P_\pm^{\alpha\beta}=
\ha(\gamma^{\alpha\beta}\pm\epsilon^{\alpha\beta})$
where $\gamma^{\alpha\beta}$ is the worldsheet metric with $\det{\gamma}=-1$
and $\epsilon^{01}=1$.    Worldsheet indices are lowered and raised
with the two-dimensional metric.
Let $\f$ denote the Grassmann envelope of
the superalgebra $\mathfrak{su}(2,2\vert 4)$, namely the Lie algebra
\begin{equation*}
\f={\cal G}r^{[0]}\otimes \mathfrak{su}(2,2\vert 4)^{[0]}\oplus {\cal G}r^{[1]}\otimes \mathfrak{su}(2,2\vert 4)^{[1]},
\end{equation*}
where ${\cal G}r$ is a real Grassmann algebra.
Introduce the two-dimensional field $g(\sigma,\tau)$ taking value in the Lie group $F$ with Lie algebra $\f$. The corresponding vector current $A_\alpha=g^{-1}\partial_\alpha g$ belongs to $\f$. The integrability
of the $AdS_5 \times S^5$ superstring action relies heavily on the
existence of an order 4 automorphism which induces a $\mathbb{Z}_4$-grading of the
superalgebra $\mathfrak{su}(2,2\vert 4)$, and thus of $\f$.
We denote by $\f^{(i)}$ the subspace of $\f$ with grade $i = 0,\cdots, 3$.
The projector on $\f^{(i)}$ shall be
denoted by  $P_i$ and we also write $M^{(i)} = P_i M$ for the projection
of $M\in\f$ on $\f^{(i)}$.
The invariant part $\f^{(0)}$ is the Lie algebra $\mathfrak{so}(4,1) \oplus
\mathfrak{so}(5)$, and the corresponding Lie group is $G=SO(4,1) \times SO(5)$.
The supertrace is compatible with the $\mathbb{Z}_4$-grading, which
means that $\str(M^{(m)} N^{(n)}) = 0$ for $m+n\neq 0$ mod $4$.

The extra ingredient needed to specify the deformation is a skew-symmetric (non-split) solution of the modified classical Yang-Baxter equation on $\f$. Specifically, this is an $\mathbb{R}$-linear operator $R$ such that, for $M, N \in \f$,
\beq
[RM, RN] - R\bigl( [RM, N] + [M, RN] \bigr) = [M, N] \label{cmYBE}
\eeq
and $\str\bigl( M RN\bigr)=-\str\bigl( RM N\bigr)$.
We shall take $R$ to be the restriction to $\mathfrak{su}(2,2\vert 4)$ of the operator
acting on the complexified algebra by $-i$ on generators associated with positive roots,
$+i$ on generators associated with negative roots, and $0$ on Cartan generators.
We will make use of the operator
$R_g=\mbox{Ad}_{g}^{-1}\circ R\circ\mbox{Ad}_g$ which is also a skew-symmetric solution of \eqref{cmYBE}.
Finally, we define the following linear combinations of the projectors,
\beqz
\dd=P_1+\frac{2}{1-\eta^2}P_2-P_3, \quad
\dt =-P_1+\frac{2}{1-\eta^2}P_2+P_3.
\eeqz
The operator $\dt$ is the  transpose operator of $\dd$ and thus
satisfies
$\str\bigl( M \, \dd(N)\bigr)=\str\bigl(\dt(M)N\bigr)$. The real
variable $\eta \in [0,1[$ will play the role of the deformation parameter.

\section{Deformed action}

As pointed out in the introduction, we will restrict ourselves in this letter to
presenting the deformed action and summarising its
most important properties. In this section we shall write down this
action and indicate the properties it shares with the
undeformed action. Properties which depend on the deformation
parameter $\eta$ are presented in the next section.

\subsection{Action}

The action, which can be obtained by generalising the method
developed in \cite{arxiv-1308.3581} to the case at hand, reads
$S[g]=\int d\sigma d\tau \Lag$ with
\beq
\Lag=- \frac{(1+\eta^2)^2}{2(1-\eta^2)} P_-^{\alpha\beta}\str\Bigl(\AA_\alpha
\, \dd\circ\frac{1}{1-\eta R_g\circ \dd}(\AA_\beta)\Bigr).
\label{def_act}
\eeq
The operator $1-\eta R_g\circ \dd$ is invertible on $\f$ for
all values of the deformation parameter $\eta \in [0, 1[$.
As in the undeformed case, there is an abelian
gauge invariance $g(\sigma,\tau) \rightarrow g(\sigma,\tau) e^{i\theta(\sigma,\tau)}$
under which the vector field $\AA_\alpha$ transforms as
$\AA_\alpha\rightarrow \AA_\alpha + \partial_\alpha\theta \, . \, \Id$. Indeed,
this leaves the action associated with \eqref{def_act} invariant because
$\str( \Id . M)=0$ for any $M$ in $\mathfrak{su}(2,2|4)$. This
invariance means that physical degrees of freedom do not
belong to the whole group $F$ but rather to the projective group $PF$.
From now on the commutators that will appear should be considered as commutators of
the projective algebra  $\mathfrak{p}\f$, and the adjoint action of $g$,
 $\mbox{Ad}_g$, is that of the projective group $PF$.
 This peculiarity already appears
in the undeformed case and the reader is referred for instance
to the review \cite{Arutyunov:2009ga} for more details.

\subsection{Original Metsaev-Tseytlin action}

The undeformed action corresponds to $\eta = 0$. Indeed, when $\eta$
vanishes, the
Lagrangian \eqref{def_act} simply becomes
\begin{align}
\Lag|_{\eta=0} &= -\ha P_-^{\alpha\beta} \str\Bigl(\AA_\alpha d|_{\eta=0} (A_\beta)\Bigr),
\nn\\
&= -\ha \str\Bigl(  \gamma^{\alpha\beta}  \AA_\alpha^{(2)}
\AA_\beta^{(2)} + \epsilon^{\alpha\beta}   \AA_\alpha^{(1)}
\AA_\beta^{(3)} \Bigr). \nn
\end{align}
One therefore recovers at $\eta=0$ the type IIB superstring action
on the $AdS_5 \times S^5$ background. This celebrated
Metsaev-Tseytlin action \cite{Metsaev:1998it} is that of a
$\sigma$-model on the semi-symmetric space $PSU(2,2|4)/G$ with
Wess-Zumino term (see for instance the reviews
\cite{Arutyunov:2009ga,Magro:2010jx}) \begin{footnote}{To be precise, note that
one must take the universal cover of $PSU(2,2|4)$
(see e.g. \cite{Beisert:2010kp}).}\end{footnote}.

\subsection{$SO(4,1) \times SO(5)$ gauge invariance}

The action corresponding to \eqref{def_act} has a gauge invariance
$g(\sigma,\tau) \rightarrow
g(\sigma,\tau) h(\sigma,\tau)$ where the function $h(\sigma,\tau)$ takes values in the subgroup $G$. This can be easily shown using the corresponding transformations
\begin{align*}
\AA_\alpha &\rightarrow h^{-1} \partial_\alpha h + \mbox{Ad}_h^{-1}(\AA_\alpha),\\
\dd(\AA_\alpha) & \rightarrow \mbox{Ad}_h^{-1} \circ \dd(\AA_\alpha),\\
R_g &\rightarrow \mbox{Ad}_h^{-1 }\circ R_g \circ \mbox{Ad}_h.
\end{align*}
This gauge transformation does not depend on the deformation
parameter $\eta$.

\section{Properties of the deformed action}

To present the properties of the action \eqref{def_act}, we will
follow the approach presented in the review \cite{Arutyunov:2009ga}
for the undeformed case.

\subsection{Equations of motion}

The equations of motion are most conveniently written in terms of the vectors
\begin{align*}
\JJ_\alpha &=  \frac{1}{1-\eta R_g \circ \dd}(\AA_\alpha), \\
\Jt_{\alpha} &=
\frac{1}{1+ \eta R_g \circ \dt}(\AA_\alpha)
\end{align*}
and their projections, $\JJ^\alpha_- = P_-^{\alpha\beta}\JJ_\beta$
and $ \Jt_+^\alpha  = P_+^{\alpha\beta}\Jt_\beta$.
In the following, we shall often use the fact that the components
$\JJ_-^0$ and $\JJ_-^1$ are proportional to each other.  One has in particular
$[\JJ_-^\alpha,\JJ_-^\beta]=0$ (and similarly for $\Jt_+^\alpha$).
The equations of motion arising from the Lagrangian
\eqref{def_act} are given by $\eom =0$ where
\begin{multline*}
\eom \coloneqq d(\partial_\alpha \JJ_-^\alpha)+
\dt(\partial_\alpha  \Jt_+^\alpha)
+\ [ \Jt_{+\alpha},
\dd(\JJ_-^\alpha)]+[\JJ_{-\alpha}, \dt( \Jt_+^\alpha)].\label{eom-original}
\end{multline*}
It is easy to check that the projection $\eom^{(0)}$ of $\eom$
onto $\f^{(0)}$ vanishes, in accordance with the gauge invariance
of the action described above.

\subsection{Rewriting the Maurer-Cartan equation}

We now wish to address the question of integrability of the
theory defined by \eqref{def_act}. Recall that in the undeformed case,
in deriving the Lax connection one makes use of the Maurer-Cartan equation $\mc=0$ satisfied by
$\AA_\alpha$, where
\beqz
\mc \coloneqq \ha \epsilon^{\alpha\beta}(\partial_\alpha
\AA_\beta-\partial_\beta \AA_\alpha+[\AA_\alpha,\AA_\beta]). \label{mcorig}
\eeqz
To find a Lax connection we therefore start by rewriting $\mc$
in terms of $\JJ_-^\alpha$ and $\Jt_+^\alpha$.
The resulting expression is a quadratic polynomial
in $\eta$. Using equation \eqref{cmYBE} for the operator $R_g$, one
can rewrite the coefficient of $\eta^2$ of this polynomial to obtain
\beqz
\mc=\partial_\alpha \Jt_+^\alpha - \partial_\alpha \JJ_-^\alpha
+[ \JJ_{-\alpha}, \Jt_+^\alpha]
+ \eta^2 [\dd( \JJ_{-\alpha}), \dt(\Jt_+^\alpha)]+\eta R_g(\eom).\label{zce}
\eeqz
Anticipating the result, let us note here that choosing $R$ to
be a non-split solution of the modified classical Yang-Baxter equation is
essential in order to preserve integrability as we deform the theory.
Before constructing the Lax connection, let us remark that the field
equations in the odd sector $P_{1,3} (\eom) = 0$ may be greatly simplified by considering the combinations
\begin{subequations}\label{idfork}
\begin{align}
P_1\circ(1-\eta R_g)(\eom)+P_1(\mc) &=-4[\Jt_{+\alpha}^{(2)}, \JJ_-^{\alpha (3)}],\\
P_3\circ(1+\eta R_g)(\eom)-P_3(\mc) &=-4[ \JJ_{-\alpha}^{(2)}, \Jt_+^{\alpha(1)}].
\end{align}
\end{subequations}
As a consequence, one can take as field equations in the odd sector
\begin{equation*}
[\Jt_{+\alpha}^{(2)}, \JJ_-^{\alpha(3)}] =0,
\qquad
[\JJ_{-\alpha}^{(2)}, \Jt_+^{\alpha(1)}] =0,
\end{equation*}
which have the same form as those of the undeformed model written in terms of ordinary currents.

\subsection{Lax connection}

We define the two vectors
\begin{multline*}
L_+^\alpha = \Jt_+^{\alpha(0)}
+ \lambda \sqrt{1+\eta^2} \Jt_+^{\alpha (1)}
+ \lambda^{-2} \frac{1+\eta^2}{1-\eta^2} \Jt_+^{\alpha(2)} \\
+ \lambda^{-1} \sqrt{1+\eta^2} \Jt_+^{\alpha(3)},
\end{multline*}
\begin{multline*}
M_-^\alpha = \JJ_-^{\alpha(0)}
+ \lambda \sqrt{1+\eta^2} \JJ_-^{\alpha (1)}
+ \lambda^{2} \frac{1+\eta^2}{1-\eta^2} \JJ_-^{\alpha(2)} \\
+ \lambda^{-1} \sqrt{1+\eta^2} \JJ_-^{\alpha(3)},
\end{multline*}
where $\lambda$ is the spectral parameter. Then, the whole set of
equations of motion $\eom=0$ and zero curvature equations $\mc=0$
are equivalent to
\beq
\partial_\alpha L_+^\alpha-\partial_\alpha M_-^\alpha+
[M_{-\alpha},L_+^\alpha]=0.\label{lax1}
\eeq
One may define an unconstrained vector
\beqz
\Lax_\alpha=L_{+\alpha}+M_{-\alpha},
 \eeqz
in terms of which the equation \eqref{lax1} becomes an ordinary zero curvature equation
\beqz
\partial_\alpha \Lax_\beta-\partial_\beta \Lax_\alpha
+[\Lax_\alpha,\Lax_\beta]=0.
\eeqz
The existence of this Lax connection shows that the dynamics of the deformed action admits an infinite number of conserved quantities.

\subsection{Virasoro constraints}

It is clear that each term in the Lagrangian \eqref{def_act} is proportional either to
the metric $\gamma^{\alpha\beta}$ or to $\epsilon^{\alpha\beta}$. The part of the
action proportional to the metric takes the form
\begin{subequations} \label{sgaga}
 \begin{align}
S_\gamma &= - \ha \left( \frac{1+\eta^2}{1-\eta^2}\right)^2
 \int d\sigma d\tau \gamma^{\alpha \beta} \str\bigl( \JJ_\alpha^{(2)} \JJ_\beta^{(2)}
\bigr),\\
&=  - \ha \left( \frac{1+\eta^2}{1-\eta^2}\right)^2
 \int d\sigma d\tau \gamma^{\alpha \beta} \str\bigl( \Jt_\alpha^{(2)} \Jt_\beta^{(2)}
\bigr).
\end{align}
\end{subequations}
To obtain this result, the skew-symmetry of $R_g$ has been used.
The Virasoro constraints are then found to be
\beqz
\str\bigl(   \Jt_{+}^{\alpha(2)} \Jt_+^{\beta(2)} \bigr) \approx 0,
\qquad
\str\bigl(   \JJ_{-}^{\alpha(2)} \JJ_-^{\beta(2)} \bigr) \approx 0.
\eeqz

\subsection{Kappa symmetry}

The invariance under $\kappa$-symmetry is a characteristic of the Green-Schwarz
formulation. We now want to show that the kappa invariance is essentially unchanged after deformation.
To do this, consider an infinitesimal right translation of the field, $\delta g=g\epsilon$,
where the parameter $\epsilon$ takes the form
\beqz
\epsilon=(1-\eta R_g)\rho^{(1)}+(1+\eta R_g)\rho^{(3)}.
\eeqz
The fields $\rho^{(1)}$ and $\rho^{(3)}$, whose expressions will be
determined shortly, respectively take values in $\f^{(1)}$ and $\f^{(3)}$. Then the
variation of the action with respect to $g$ reads
\begin{multline*}
\delta_g S= \mbox{$\frac{(1+\eta^2)^2}{2(1-\eta^2)}$}\int d\sigma d\tau\str
\Bigl(\rho^{(1)}P_3\circ(1+\eta R_g)(\eom) \\
+\rho^{(3)}P_1\circ(1-\eta R_g)(\eom)\Bigr).
\end{multline*}
We may then use equations \eqref{idfork} to write this variation as
\begin{multline*}
\delta_g S = -\mbox{$2 \frac{(1+\eta^2)^2}{(1-\eta^2)}$} \int d\sigma d\tau\str\Bigl(
\rho^{(1)} [ \JJ_{-\alpha}^{(2)}, \Jt_+^{\alpha(1)}] + \\
+\rho^{(3)} [ \Jt_{+\alpha}^{(2)}, \JJ_-^{\alpha(3)}] \Bigr).
\end{multline*}
In full analogy with the undeformed case (see \cite{Arutyunov:2009ga}),
we take the following ansatz for $\rho^{(1)}$ and $\rho^{(3)}$:
\begin{align*}
\rho^{(1)} &= i \kappa^{(1)}_{+\alpha}  \JJ_-^{\alpha (2)}
 +    \JJ_-^{\alpha (2)}  i \kappa^{(1)}_{+\alpha} ,\\
\rho^{(3)} &= i \kappa_{-\alpha}^{(3)} \Jt_+^{\alpha (2)}
+   \Jt_+^{\alpha (2)}  i \kappa_{-\alpha}^{(3)},
\end{align*}
where $\kappa^{(1)}_{+}$ and $\kappa^{(3)}_{-}$ are constrained
vectors of respective gradings $1$ and $3$. Note that we are using the standard
convention for the real form $\mathfrak{su}(2,2|4)$ (see for instance appendix $C$ of
\cite{Grigoriev:2007bu}). Then, a short calculation leads to
\begin{align*}
 \str\bigl( \rho^{(1)} [\JJ_{-\alpha}^{(2)}, \Jt_+^{\alpha(1)}]\bigr)
&= \str\bigl( \JJ_-^{\alpha(2)} \JJ_-^{\beta(2)} [\Jt_{+\alpha}^{(1)}, i \kappa_{+\beta}^{(1)}]
\bigr),\\
 \str\bigl( \rho^{(3)} [\Jt_{+\alpha}^{(2)}, \JJ_-^{\alpha(3)}]\bigr)
&= \str\bigl( \Jt_+^{\alpha(2)} \Jt_+^{\beta(2)} [\JJ_{-\alpha}^{(3)}, i \kappa_{-\beta}^{(3)}]
\bigr).
\end{align*}
 At this point, we use the standard property (see \cite{Arutyunov:2009ga})
that the square of an element of grade $2$ only contains
a term proportional to $W=\mbox{diag}(\Id_4,-\Id_4)$ and a term proportional to the
identity which does not play a role in the case at hand. We finally obtain
\begin{multline*}
 \delta_g S=-\mbox{$\frac{ (1+\eta^2)^2} {4(1-\eta^2)} $}\int \!\! d\sigma d\tau \Bigl( \str(\JJ_-^{\alpha(2)}
\JJ_-^{\beta(2)}) \times\\ \times
  \str\bigl(W [\Jt_{+\alpha}^{(1)}, i \kappa_{+\beta}^{(1)}]\bigr)
+  \str(\Jt_+^{\alpha(2)}
\Jt_+^{\beta(2)}) \str\bigl( W [\JJ_{-\alpha}^{(3)}, i \kappa_{-\beta}^{(3)}]\bigr) \Bigr).
\end{multline*}
This expression comes from the variation of the field $g$ in the action.
It may be compensated by another term coming from the variation of the
metric $\gamma$. To determine this variation we use the result \eqref{sgaga}.
We are then led to choose
\beqz
\delta\gamma^{\alpha\beta}=
\frac{ 1-\eta^2}{2}  \str\Bigl(    W [ i \kappa_+^{\alpha(1)},\Jt_+^{\beta (1)} ]
+ W [i \kappa_-^{\alpha(3)},\JJ_-^{\beta(3)} ]   \Bigr)
\eeqz
for the transformation of the metric in order to ensure $\kappa$-symmetry.


 \section{Conclusion}

The Lagrangian \eqref{def_act} is a semi-symmetric space generalisation
of the one obtained in \cite{arxiv-1308.3581} by deforming
the symmetric space $\sigma$-model on $F/G$. In the latter case,
it was shown that the original $F_L$ symmetry is deformed
to a Poisson-Hopf algebra analogue of $U_q(\f)$. The same fate is confidently expected for
the $\mathfrak{psu}(2,2|4)$ symmetry of the $AdS_5 \times S^5$
superstring. Hence, the $q$-deformation proposed here
 generalizes the situation which holds for the squashed sphere $\sigma$-model
\cite{Kawaguchi:2012ve,Kawaguchi:2012gp}.

As mentioned in the introduction, the construction of the deformed theory
relies on the existence of a second compatible Poisson bracket. The latter
is known to be related \cite{Delduc:2012vq} to the Pohlmeyer reduction of
the $AdS_5 \times S^5$ superstring \cite{Grigoriev:2007bu,Mikhailov:2007xr}.
In fact, one motivation for deforming the superstring action
comes from the $q$-deformed $S$-matrix appearing in this
context \cite{Hoare:2011fj, Hoare:2011wr, Hoare:2012fc}, built from the
$q$-deformed $R$-matrix of \cite{Beisert:2008tw}.
It would therefore be very interesting to make contact between these
two deformations.

Let us end on a more conjectural note by commenting on
the limit $\eta \to 1$ of the deformed model.
The analogous limit in the case of the deformed $SU(2)/U(1)$ $\sigma$-model
corresponds to a $SU(1,1)/U(1)$ $\sigma$-model \cite{arxiv-1308.3581}.
If such a property were to generalise to the case at hand, we expect
that the cosets $AdS_5 \simeq SO(4,2)/SO(4,1)$ and $S^5 \simeq SO(6)/SO(5)$
would respectively be replaced in this limit by
$SO(5,1)/SO(4,1) \simeq dS_5$ and $SO(5,1)/SO(5) \simeq H^5$.
Such cosets have already been considered in \cite{Hull:1998vg}.
This point certainly requires closer investigation and we will come back to it
from the hamiltonian point of view elsewhere.

\end{document}